\documentclass[12pt]{article}       
\begin{document}

\title{Cosmic Accelerators  }

\author{ Jonathan Arons}

\date{February 15, 2009} 

\maketitle

\begin{center}
{\it A White Paper submitted to the Astronomy 2010 Decadal Survey\\
\hspace*{1in}
\newline
Endorsed by \\
the Topical Group in Plasma Astrophysics (TGPAP) of the American Physical Society,\\
and the NSF Physics Frontier Center for Magnetic Self-Organization (CMSO)}
\end{center}

\abstract{I discuss the scientific rationale and opportunities in the study of high energy 
particle accelerators away from the Earth; mostly, those outside the Solar System. I also
briefly outline the features to be desired in telescopes used to probe accelerators studied
by remote sensing.}
\vspace*{0.5in}

Relativistic Astrophysics generates a number of problems where plasma
physics plays an essential role.  The acceleration of cosmic rays, and
of the nonthermal particle distributions present in astrophysical sources 
of synchrotron radiation, are famous and long standing examples.  The
acceleration, transport and termination of magnetized relativistic flows 
emerging from compact objects is an almost equally hoary astrophysical problem, first recognized in the discovery of apparently super-luminal
jets emerging from active galactic nuclei in the early 70s, as well as implied by modeling of the energy losses from rapidly rotating magnetized
neutron stars (pulsars), also beginning in the late 60s-early 70s.  

Plasma physical results play a central role in the theoretical modeling of
these phenomena. Many have their roots in non-relativstic plasma studies,
often driven by space plasma physics experiments.

Experiments from satellites immersed in the solar wind drove the 
development of quasi-linear theories, and later computational models, of high energy particle diffusion in long wavelength magnetic turbulence, with particles undergoing strong scattering from waves with wavelengths comparable to the particle Larmor radius.  For cosmic rays and other high energy (supra-thermal) particles, the Larmor radii are large compared to
the thermal Larmor radii of the underlying plasma, causing the particles to
resonate with long wavelength waves - for ions, these are MHD modes, thus
coupling particle transport to the properties of MHD turbulence in the medium (the solar wind, in the space plasma experimental context.)

Outside the solar system, experimental information on MHD turbulence in the
media in which cosmic rays propagate - the interstellar medium of our and 
other galaxies, and the intergalactic medium within and between clusters - was and is largely lacking, except at the longest wavelengths where the turbulence is forced.  The fact that cosmic rays themselves can generate MHD turbulence, through a resonant cyclotron instability driven by their streaming through the medium, was a
significant discovery in the late 60s (Lerche 1967, Kulsrud and Pearce 1969), augmented by the discovery in the last 10 years that the cosmic ray electric current near a cosmic ray source can drive a powerful non-resonant instability (Bell 2004) perhaps responsible for the strong magnetic fields inferred in supernova remnants (e.g., Berezhko {\it et al.} 2003, Vink \& Laming 2003, Volk {\it et al.} 2005).  The back reaction of these waves on the cosmic rays reduces their streaming velocity to no more than the local Alfven speed in the {\it total} magnetic field, at least so long as the energy density in the relativistic component of the plasma (the cosmic rays) does not exceed the rest energy density of the main, non-relativistic constituent.  This application of theoretical plasma physics to astrophysics has had major influence on all modeling of cosmic ray transport and origins, including high energy particles in the low density interstellar medium (observed through galactic synchrotron radiation and $\pi 0$ gamma ray emission);
cosmic rays in molecular clouds ($\pi 0$ gamma ray emission); cosmic ray
electrons in galaxy clusters (synchrotron emission); etc.  

In the later 70s, the first really successful theory of cosmic ray acceleration appeared.  Prefigured by earlier work on acceleration in
the solar wind (Fisk 1971), several authors 
(Axford {\it et al.} 1977, Krymsky 1977, Bell 1978, Blandford \& Ostriker 1978 ) simultaneously identified Fermi acceleration in nonrelativistic shock waves in supernova remnants as the likely culprit behind the delivery of $\sim$ a supernova explosion's kinetic energy to relativistically high energy particles.  The diffusion of cosmic
rays in MHD turbulence was the key plasma conceptual development of this idea,
now accepted as the paradigm for high energy particle acceleration. Diffusion of particles in the media up- and down-stream of the shock provides the ``mirrors'', which force
particles to repeatedly cross the shock front, with fractional energy gain
per crossing $\sim v_{shock}/c $. Upstream turbulence is thought to arise 
from the cyclotron instability driven by the cosmic rays' diffusing ahead of the shock wave in the underlying fluid. This ``diffusive fermi acceleration'' (DFA) or ``diffusive shock acceleration'' (DSA) has swept the field, even though direct experimental and computational evidence for it actually being at work is rather limited.

Circumstantial evidence for this
process being in operation in supernova remnant shocks has come from
studies of the detailed radio synchrotron spectra of supernova 
remnants, and from X-ray studies suggesting that the shock compressions exceed those of adiabatic shocks, indicating the presence of high energy
particle populations with pressure comparable to the shock kinetic
energy density, a result in accord with predictions from models of shock
acceleration in which the scattering turbulence is supplied as a phenomenological part of the model. 

MHD turbulence theory itself seriously advanced in the 90s, with the appearance of the Goldreich-Sridhar (1995) model, the first extension of 
Kolmogorov's ideas to the MHD realm to incorporate the essential anisotropy of the magnetic stresses.  This theory has been applied to
the transport of cosmic rays in the general interstellar medium, with rather surprising results - the turbulence anisotropy greatly lengthens the mean free paths (Chandran 2000). The consequences of this discovery are still not fully
appreciated, or worked out, either for general cosmic ray transport, for
shock acceleration and for high energy particles in galaxy clusters (but see 
Brunetti \& Lazarian 2007 for an example where this issue is confronted, in the
galaxy cluster context.) 

All of these advances apply to fundamentally non-relativistic astrophysical
plasmas permeated by relativistic constituents which are small by number,
and have energy densities no more than comparable to the nonrelativistic
pressures and flow energy densities. The discovery of pulsars (Hewish {\it et al.} 1968), with their implication of fully relativistic plasmas and relativistically strong 
magnetic fields, opened up a very different regime of relativistic plasma 
astrophysics.  The realization that jets from active galactic nuclei probably are also fully relativistic opened the door into the same plasma world. Relativistic magnetospheres of neutron stars and their winds, and relativistic disks around black holes and {\em their} winds and jets, drives the development of relativistic MHD  and its force free subset (Gruzinov 1999, 2005), to the point where now practitioners of astrophysical MHD probably know more about transalfvenic magnetized flow than is found in any of the lab-based areas of plasma physics. Relativistic MHD differs from the more familiar 
nonrelativistic version in its elevation of electric stresses to 
significance comparable to the magnetic stresses, as well as raising new
problems in magnetic dissipation and reconnection.  

Especially in the almost 20 years 1990-2009, advances in computational technique have allowed the development of dynamical models of magnetized relativistic flow which open the door to quantitative modeling of the observations of these exotic 
systems ({\it e.g.} Gammie {\it et al.} 2003). In parallel, simplified analytic models used both to interpret the data and the simulations of ideal MHD flow have greatly advanced - theories of the relativistic winds from pulsars play a central part in modeling pulsar wind nebulae; theories of relativstic jet flow are prominent in the interpretation of the sporadic jets observed in microquasars - $10 M_\odot$ black holes in binary star systems - in the formation and propagation of relativistic jets in Blazars and Gamma Ray Burst Sources.

The disk-jet connection is a fundamental challenge to relativistic MHD flow
theory, as applied to the formation of jets from the disks around black holes. Recent radio, infrared and X-ray observations open the possibility of observationally constrained modeling of jet formation, a fundamentally multi-dimensional challenge to computational modeling of accretion disks.  In particular, an understanding of the circumstances under which disks form a macroscopic magnetic field, in addition to the small scale turbulence generated by the Magnetorotational Instability (Balbus and Hawley 1991, 1998), is sorely lacking.  Likewise, two and three dimensional relativistic MHD models of the formation of winds from rotating magnetospheres have just begun.  Both these areas pose severe challenges to model building, since the relativistic motion and electromagnetic stress dominance create formidable problems to existing computational technique (the equations of motion become intractably stiff), as well as largely escaping analytic methodologies.

Furthermore, such modeling requires incorporation of magnetic dissipation,
usually at current sheets. Here, basic questions of resistive behavior in relativistic plasmas, and associated issues of the local dynamics of reconnection, become central.
Current methodology mostly relies upon replacing the physics of resistivity with the dissipation inherent in the numerical algorithms - exceptions ({\it e.g.} Komissarov  2007) use a resistive MHD model with an isotropic resistivity specified as a parameter. Experience with non-relativistic current sheets measured in space, especially in the Earth's magnetosphere, suggest that such approximations and models are often not a proper representation even of the qualitative nature of the physics. The understanding of reconnection in the relativistic environment needed to improve on this state of affairs has hardly begun; the development of such understanding, through theory and kinetic simulation  as well as the incorporation of that understanding into macroscopic flow models, is a crucial requirement for advancing the systematic modeling of these relativistic environments, for which laboratory  experiments are unlikely.  

Reconnection at current sheets in highly magnetized ($\sigma = B^2/4\pi \rho \gamma c^2 \gg 1$) flows
is often mentioned as a mechanism of relativistic particle acceleration and consequent synchrotron emission. Our understanding of this is primitive,
yet crucial for connecting the observations to the system models in a quantitative (or even a qualitative!) manner.  In systems where the magnetization is not as large, the dissipation, particle acceleration and resulting photon emission is usually ascribed to acceleration at relativistic shock waves - these are inefficient in high $\sigma$ flows. Substantial development has gone into migrating the DSA mechanism to the relativistic environment. As in the nonrelativistic case, the  turbulence needed to  scatter particles back and forth across the shock front has been {\it assumed} to be present. Monte Carlo simulations have shown that large amplitude magnetic turbulence is required to provide sufficient scattering - the high flow velocities lead to escape of particles from the shock before much energy gain occurs, unless the turbulence is very strong.  

In the last two decades, particle-in-cell simulation techniques have been applied to the
relativistic shock problem, for shocks with and without substantial upstream macroscopic magnetic field.  Relativistic shock simulations have just begun to show solid evidence for nonthermal particle acceleration, including evidence for the high turbulence levels required in the DSA phenomenological models.  The clearest evidence for this and other
nonthermal acceleration mechanisms in relativistic shocks comes from studies of shocks in electron-positron plasmas (Spitkovsky 2008).  On the other hand, some evidence for non-DSA acceleration exists from 1D simulations of shocks  where higher energy particles (``protons'') carry most of the energy flux (Amato \& Arons 2006).  Deeper resolution of these issues awaits the rapidly improving ability to do 3D simulations. Such progress is essential, if relativistic collisionless shock heating and particle acceleration is to become usable as part of the process of modeling relativistic astrophysical flows.\\

These various achievements and opportunities illustrate two lessons.  \\

1) Astrophysical plasma physics begins in the modeling of astronomical (including solar system) observations, pursued by plasma informed modelers. This history suggests the great progress will come from better plasma education for astrophysicists.  The greatest advances of all are likely to come from plasma educated astrophysicists teaming up with plasma physicists coming out of laboratory and space plasma physics backgrounds. The former have the system modeling instincts, combined with sufficient plasma physics involvement, to allow ready communication across the astrophysics-plasma physics barrier.  The latter have skills and techniques which can allow much more rapid progress than is possible when the astrophysicists have to reinvent all the wheels.  The glue is provided by computational plasma physics applied to astrophysical problems, since astrophysical impact requires taking  the nonlinearities inherent in the systems' physics .  The last 15 years' advances in sophisticated simulation methods, and the computer horsepower to use them in reasonably realistic manners, combined with the tremendous  advances in observations of relativistic astrophysical plasmas, foretell a bright future for this marriage of disciplines. 

2) The subject is rooted in observation, as is all of astrophysics.  Cosmic accelerators manifest themselves in the presence of high energy particles, observed directly (cosmic rays, both fast particles of solar system origin and high energy particles from outside the solar system), and by observation of non-thermal photon spectra.  Photon observations require full multi-wavelength capability, from multi-TeV energies currently detected from the ground (Cerenkov telescopes) and MeV-GeV energies detected from space, reflecting synchrotron and inverse Compton emission of high energy electrons (and positrons) and $\pi$0 decays from hadronic interactions, down to micro-eV radio photons which reflect synchrotron emission.  The study of cosmic accelerators thus relies upon continually improving the sensitivity of the full range of astronomical facilities.
 
Acceleration occurs because of dissipative structures in the plasmas (shocks, current sheets, turbulence), whose physical structure is small in at least one dimension compared to the size of the overall system.  Therefore, high angular resolution and concomitant high sensitivity at all photon energies are essential for this field, while high spectral resolution is generally of less significance. The full suite of advanced telescopes now available and under development for the next decade are essential for improving our understanding of cosmic accelerators. The recently launched space based FERMI gamma ray observatory (Attwood {\it et al.} 2009) plus the existing (H.E.S.S., Cangaroo, VERITAS) and forthcoming (CTA, HAWC) ground based atmospheric and water Cerenkov telescopes provide essential input into the acceleration physics.  Focusing X-ray missions such as Simbol-X and the development of Laue lens technology for
soft gamma ray imaging will be of great use, since such experiments directly sample the high energy electron (and positron) population in relativistic sources (as do the high energy gamma ray telescopes); the focusing instruments allow distinguishing the dissipative structures where particles are accelerated from the larger scale regions where the particles do most of their emission.  Since radiation losses are rapid at high energy, X- and $\gamma$-ray measurements provide the best probes of contemporary particle acceleration; short radiative lifetimes also allow one to probe variability in the accelerators, itself a strong probe of the physics. Radio measurements of synchrotron emission at high angular resolution are especially useful in constraining the overall particle population - since synchrotron emission is less efficient at low energy, these observations provide significant constraints on the rest mass budget of the accelerators. Infrared observations are especially useful in linking the radio to the high energy photon regimes, thus allowing one to unravel the effects of radiative cooling on the nonthermal particle population from energy space structure in the accelerators themselves, as has been the case in the recent use of Spitzer observations of a pulsar wind nebula (Slane {\it et al.} 2008).

Direct laboratory experiments on relativistic plasmas are difficult, although the onset of experiments in which petawatt laser irradiation of solid foil targets leading to production of electron-positron plasmas ({\it e.g.} Wilks {\it et al.} 2005, Remington 2006) is an exciting advance which will provide significant input into the study of relativistic cosmic accelerators.  Experiments on non-relativistic magnetic reconnection, such as MRX ({\it e.g.} Yamada {\it et al.} 1997, Ren {\it et al.} 2008), whose dynamics in the collisionless regime is analogous to what is expected to be the case for relativistic collisionless reconnection, certainly will advance our understanding of the physics.  Experiments on the dissipation regime of MHD turbulence ({\it e.g.} Carter 2006) can be informative on the turbulence expected to be present in non-relativistic shock waves acting as particle accelerators.  

Much to be desired would be the development of an experiment to observe shocks with scale large enough to observe the relatively slow processes in shock acceleration.  Such experiments actually are best done in the solar wind. Interestingly, the laser-plasma experiments mentioned above have a decent chance of eventually observing such phenomena in the laboratory, through creation of large scale colliding, cooled electron-positron jets with small skin depth lengths less than the size of the plasmas.

Experiments in the solar wind and in the Earth's magnetosphere provide fundamental input into our understanding of current sheets and shocks. Even though these systems are not perfect models of the astrophysical environment, they have the advantage over laboratory experiments of being systems which are much larger than the microscopic plasma lengths scales, thus allowing investigation of the phenomena without interference from interactions with walls that are not present in the astrophysical environment. 

Computation and theory are the other legs of astrophysical plasma investigations.  Improvements in computer capability and availability are common necessities in all branches of astrophysics.  More important than machines are the algorithmic developments needed to take advantage of petascale (and exascale, in principle) computation.  Building new computational tools is now sufficiently challenging that new modes of research, in which teams of plasma astrophysicists, applied mathematicians and computer scientists will have to work together, to build new codes and new data analysis methodologies to extract the useful information.  ``Pure'' theory will also continue to have an important role, as a means of motivating and interpreting computations, as well as directly interpreting observational data. \\

\noindent {\Large \bf References}  \\

\noindent Amato, E., \& Arons, J. 2006, Ap.J., 653, 325

\noindent Attwood, W.B., {\it et al.} 2009, submitted to Ap.J. (arXiv: 0902.1089)

\noindent Axford, W.I.,  {\it et al.} 1977, Proc. 15th Int. Cosmic Ray Conf., 11, 132

\noindent Balbus, S.A., and Hawley, J.F.  1991, Ap.J., 376, 214; 1998, Rev. Mod. Phys., 70, 1

\noindent Bell, A.R. 1978, MNRAS, 182, 147

\noindent Bell, A.R. 2004, MNRAS, 353, 550

\noindent Berezhko, E. G., Ksenofontov, L. T., \& Volk, H. J. 2003, A\&A, 412, L11

\noindent Blandford, R., \& Ostriker, J. 1978, Ap.J., 221, L29

\noindent Brunetti, G., \& Lazarian, A.\ 2007, MNRAS,  378, 245

\noindent Carter, T.A. 2006, Phys. Plasmas, 13, 010701 

\noindent Chandran, B., 2000, Phys. Rev. Lett., 85, 4656

\noindent Fisk, L.A. 1971, J. Geophys. Res., 76, 1662

\noindent Gammie, C.F., McKinney, J.C., and Toth, G. 2003, Ap.J., 589, 444

\noindent Goldreich, P., and Sridhar, S. 1995, Ap.J., 438, 763

\noindent Gruzinov, A. 1999, astro-ph/9902288

\noindent Gruzinov, A. 2005, PRL, 94, 021101

\noindent Hewish, A., Bell, S.J., Pilkington, J.D.,  {\it et al.} 1968, Nature, 217, 709

\noindent Komissarov, S. 2007, MNRAS, 382, 995

\noindent Krymsky, G.F. 1977, Sov. Phys. Dokl., 23, 327

\noindent Kulsrud, R.,  and Pearce, W.M.1969, Ap.J., 156, 445

\noindent Lerche, I. 1967, Ap.J., 147, 689

\noindent Remington, B.A., Drake, R.P., and Ryutov, D.D. 2006, Rev. Mod. Phys., 78, 755

\noindent Ren, Y., Yamada, M.,  Ji,  H., {\it et al.} 2008, Phys. Plasmas, 15, 082113

\noindent Slane, P., Helfand, D., Reynolds, S.P., {\it et al.} 2008, Ap.J., 676, L33

\noindent Spitkovsky, A. 2008, Ap.J., 682, L5

\noindent Vink, J., \& Laming, J.M. 2003, Ap.J., 584, 433, 229 

\noindent Volk, H.J., Berezhko, E.G., and Ksenofontov, L.T.  2005, A\&A, 409, 563

\noindent Wilks, S.C. {\it et al.} 2005, Ap. and Space Sci., 298, 347

\noindent Yamada, M., Ji, H., Hsu, S., {\it et al.} 1997, Phys. Plasmas, 4, 1936

\end{document}